\documentstyle[prl,aps]{revtex}
\begin{document}

\draft
\title{Orthorhombically Mixed s and d$_{x^2-y^2}$ Wave
Superconductivity\\
and Josephson Tunneling}

\author{M.B. Walker}

\address{Department of Physics\\
University of Toronto\\
Toronto, Ontario, M5S 1A7\\
Canada\\}

\date{June 26, 1995}

\maketitle
\begin{abstract}

The effect of orthorhombicity on Josephson tunneling in
high T$_c$ superconductors such as YBCO is studied
for both single crystals and highly twinned crystals.
It is shown that experiments on highly twinned
crystals experimentally
determine the symmetry of the superconducting twin
boundaries (which can be either even or odd with
respect to a reflection in the twinning
plane).  Conversely, Josephson experiments on
highly twinned crystals can not experimentally determine whether
the superconductivity is predominantly $s$-wave or
predominantly $d$-wave. The direct experimental determination
of the order-parameter symmetry by Josephson tunneling in
YBCO thus comes from the relatively few experiments which
have been carried out on untwinned single crystals.

\end{abstract}
\twocolumn
\newpage

{\it Introduction.}
Although many of the currently
studied high-temperature
superconductors are orthorhombic (including
YBa$_2$Cu$_3$O$_{7-\delta}$
(YBCO) and Bi$_2$Sr$_2$CaCu$_2$O$_{8+\delta}$,
\protect\cite{Haz90}) they are often regarded
as being
approximately tetragonal. Thus, in the current
debate concerning
the symmetry of the order parameter, the
attention has been
focussed on attempting to distinguish between
order parameters
having $s$, $d_{x^2-y^2}$ or possibly
other symmetries (see \protect\cite{Dyn94}
for a recent review).
As a result of the orthorhombicity of the high T$_c$
superconductors, however, the $s$ and $d_{x^2-y^2}$ order
parameters are coupled \protect\cite{Li93,Odo95}.
This note discusses some of the
consequences of orthorhombicity and of this coupling
in the interpretation of Josephson tunneling experiments,
and focuses on what can be directly measured experimentally
without relying on microscopic theories or models of the
superconducting state and the twin boundaries.

Experimental studies of the Josephson effect have been
suggested \protect\cite{Ges86,Sig92}
as a method of establishing the order parameter
symmetry in the case of unconventional superconductivity.
Measurements of the basal-plane Josephson currents in
YBCO \protect\cite{wol93,Mat95,Igu94,Bra94,Tsu94}
have found these currents to be characteristic of
$d$-wave symmetry.  On the other hand, a measurement
\protect\cite{Sun94} of the $c$-axis
Josephson current raised questions
about the $d$-wave interpretation.

Below, I study the Josephson currents from an
orthorhombic superconductor such as YBCO to an $s$-wave
superconductor such as Pb for both single crystal and
highly twinned (all twins occurring with equal weight)
samples.  The superconducting state of YBCO can be
shown to be either even or odd with respect to reflection in
a twinning plane, and it is demonstrated
that Josephson tunneling experiments on highly
twinned samples determine this symmetry.

Contrary to current views, however, Josephson
experiments on highly twinned samples can not determine
whether the superconductivity is predominantly $s$-
or $d$-wave.  The Josephson current from a single
twin contains both $s$- and $d$-wave components (as it
must because a single twin is orthorhombic).  The
averaging over all twins, however, effectively imposes
tetragonal symmetry and results in a Josephson
current which depends on only a single symmetry component
(either $s$- or $d$-wave); which component is selected
depends on the superconducting twin-boundary symmetry.
Josephson experiments on highly twinned
samples will thus measure the magnitude of the selected component,
but will give no information on the magnitude of the
other component.  In order to determine whether the
symmetry of the order parameter is predominately $d$-wave
or predominately $s$-wave, it is essential to measure the
relative magnitude of the two components (which can be done
on untwinned single crystals, but not on twinned crystals).

In view of the above, the direct experimental determination
of the order-parameter symmetry by Josephson tunneling in
YBCO thus comes from the relatively few experiments
\protect\cite{Bra94,wol93} which
have been carried out on untwinned single crystals.

{\it Mixed $s$- and $d$-wave superconductivity.}
The Ginzberg-Landau free energy density describing coupled $s$-wave
and $d_{x^2-y^2}$\ order parameters ($\psi_s$ and $\psi_d$,
respectively) is
\begin{eqnarray}
F&=&\alpha_s\vert\psi_s\vert^2 +\alpha_d\vert\psi_d\vert^2
\nonumber\\
&+&\frac{1}{2}\beta_s\vert\psi_s\vert^4 +\frac{1}{2}\beta_d\vert
\psi_d\vert^4+\beta_4\vert\psi_s\vert^2\vert\psi_d\vert^2
\nonumber\\
&+& (-1)^\epsilon \alpha(\psi_s\psi_d^\ast +\psi^\ast_s\psi_d)
+\beta_0(\psi_s^2\psi_d^{\ast 2}+\psi_s^{\ast 2}\psi_d^2)\nonumber\\
&+&(-1)^\epsilon
[\beta_1\vert\psi_d\vert^2+\beta_2\vert\psi_s\vert^2](\psi_s\psi_
d^\ast+\psi_s^\ast\psi_d) \nonumber
\end{eqnarray}
here $\epsilon=1,2$ refers to one of the two possible twin
orientations which occurs in the YBCO materials.
Except for the factors of $(-1)^{\epsilon}$, which are
important below, this free energy has been given in
\protect\cite{Li93}.  I assume
that the twin boundary between two twins (as in Fig. 1)
is a plane of reflection symmetry of the underlying
crystal lattice.  If the $d$- and $s$-wave
order parameters describing superconductivity in twin 1
are $(\psi_d, \psi_s)$, then a reflection in the
twinning plane gives a superconducting state of twin 2
described by order parameters $(-\psi_d, \psi_s)$
and having the same free energy as the corresponding state of
twin 1; hence the factor $(-1)^{\epsilon}$ which appears at
various places in the free energy.

By definition, $d$-wave superconductivity will be modelled by the
assumption that $\alpha_d =\alpha_d^\prime(T-T_{d0})$, while
$\alpha_s$ is taken to be positive. (Conversely, $s$-wave
superconductivity will be modelled by taking $\alpha_s
=\alpha_s^\prime (T-T_{cs})$ with $\alpha_d > 0$.) For $d$-wave
superconductivity, minimizing $F$ with respect to $\psi^\ast_s$ in
zero external magnetic field gives $\psi_s$ correct to terms linear
in $\psi_d$ as
$$
\psi_s =(-1)^{\epsilon+1}(\alpha/\alpha_s)\psi_d
$$
Also $\vert \psi_d\vert ^2 =\alpha_d^\prime(T_d -
T)/\tilde{\beta}_d$; here $\tilde{\beta}_d$ is a constant
approximately equal to $\beta_d$ if $\alpha$ is small, and $T_d$
differs from $T_{d0}$ when $\alpha$ is non zero.

Since $\psi_s$ can be found as an expansion in powers of
$\psi_d$ (and this is true for both $s$- and $d$-wave
superconductivity as these have been defined above),
$\psi_s$ can be substituted for in terms of $\psi_d$
only (the coupling to $\psi_s$ nevertheless being
implicitly taken into account).

The nature of the superconducting twin boundary between
twins 1 and 2 will be determined by the interface free
energy per unit area
$$
F_{12} = B(\psi_{d1}^{\ast}\psi_{d2} + \psi_{d1}\psi_{d2}^{\ast}),
$$
where, as just explained, $\psi_s$ is implicitly taken
into account.  If $B<0$ (or $B>0$), the free energy
is minimized by taking $\psi_{d1} = \psi_{d2}$
(or $\psi_{d1} = -\psi_{d2}$), and the superconducting
state will be odd (or even) under reflection in the twin
boundary, respectively.  Note that if the twin boundary
is odd, $\psi_{d1} = \psi_{d2}$, but that because of the
factor $(-1)^{\epsilon}$ in the above free energy,
$\psi_{s1} = -\psi_{s2}$.  Thus, calling a twin boundary
a normal or a $\pi$ boundary according
to whether or not the order parameter changes sign could be
ambiguous, and it is better to identify the type of twin
boundary by its reflection symmetry.

{\it Basal-plane Josephson currents.}
The twin boundaries in YBCO are normal to the [110]
and [$\,\overline 1$10] directions.  A YBCO crystal will
consist of some regions where orthorhombic twins are
separated by [110] twin boundaries, and other regions
where orthorhombic twins are separated by [$\,\overline 1$10]
twin boundaries.  For want of a better name, I will use the word
``macrodomain'' to refer to a region where the twin boundaries all
have the same orientation.  Thus twinned YBCO can be considered to
be made up of [110] macrodomains and [$\,\overline 1$10]
macrodomains. I begin by discussing a
[$\,\overline 1$10]  macrodomain,
which contains [$\,\overline 1$10] twin boundaries and two types
of twins (twin 1 and twin 2) as shown in Fig.1.

Consider a Josephson
junction between twin 1 of YBCO and Pb.  The normal {\bf n}
(directed from YBCO to Pb) to the plane of the junction
is taken to lie in the YBCO a-b plane and to make an angle
$\theta$ with the a-axis of twin 1 (see Fig.1).
The Josephson current from twin 1 to Pb is
$$
j_1(\theta) = i g(\theta)(\psi_{d1}^{\ast}\psi_{Pb}
- \psi_{d1}\psi_{Pb}^{\ast}).
$$
Now write
$$
g(\theta) = d(\theta) + s(\theta)
$$
where
$$
d(\theta) = [g(\theta) - g(\theta + \pi /2)]/2,
$$
$$
s(\theta) = [g(\theta) + g(\theta + \pi /2)]/2.
$$
The constraints of orthorhombic symmetry require
$$
g(\theta) = g(-\theta),\ g(\theta + \pi) = g(\theta),
$$
$$
d(\theta +\pi /2) = -d(\theta),\
s(\theta +\pi /2) = s(\theta).
$$
It is clear from these symmetry properties of $d(\theta)$
and $s(\theta)$ (and this will appear in more detail
below) that these functions will exhibit characteristic
$d_{x^2-y^2}$-wave and $s$-wave behavior, respectively, in
Josephson tunneling experiments; the
above separation of $g(\theta)$ into $d(\theta)$
and $s(\theta)$ is thus a fundamental step in the
interpretation of the basel plane Josephson tunneling
from an orthorhombic superconductor to an $s$-wave
superconductor in terms of $d$- and $s$-wave components.
The following
Fourier series representations of the functions $g(\theta)$,
$d(\theta)$ and $s(\theta)$ incorporate the
above-mentioned symmetries:
$$
g(\theta) = \Sigma\ g_n cos(2n\theta),
$$
$$
d(\theta) = \Sigma\ d_{2+4n}cos[(2+4n)\theta],
$$
$$
s(\theta) = \Sigma\ s_{4n}cos(4n\theta).
$$
In all cases the sums are over all integral n from
zero to infinity.

The Josephson current from twin 2 to Pb in the direction $\theta$
can be obtained from that for twin 1 by reflecting in the twin-twin interface
(see Fig. 1).  Hence
$$ \textstyle
j_2(\theta) = -ig({1\over 2}\pi + \delta - \theta)(\psi_{d2}^{\ast}
\psi_{Pb}-\psi_{d2} \psi_{Pb}^{\ast}).
$$

For the purpose of interpreting corner squid experiments, the
current along two orthogonal directions in the basal plane (which
will be called the $x$ and $y$ directions) will be calculated; the $x$
and $y$ directions are chosen to be at angles of $\theta = \theta_1
+\delta /2$ and $\theta = \theta_1 +(\pi + \delta)/2$ from the
{\bf a}$_1$ direction (see Fig. 1).  The angle $\theta_1$ (just defined
in terms of $\theta$ by the preceding equations) is used rather than
$\theta$ to define the $x$ and $y$ directions because $\theta_1 = 0$
is a symmetrical orientation of $x$ and $y$ relative to the two twin
orientations.  Finally, the expressions given below for the Josephson
currents will be for currents averaged with equal weights over the two
twins in a [$\,{\overline 1}$10] macrodomain, and will be given in the form
$$
j_{avg,\alpha}=c_{\alpha}(\theta_1)|\psi_d \psi_{Pb}|sin(\phi_d -\phi_{Pb})
$$
where $\alpha$ is $x$ or $y$, $\psi_d = |\psi_d|exp(i\phi_d)$, and
$\psi_{Pb} = |\psi_{Pb}|exp(i\phi_{Pb})$.

For the case of even twin-twin interfaces, $\psi_{d1}=-\psi_{d2}=\psi_d$,
and
$$
c_x(\theta_1)=+[d(\theta_1^+) - d(\theta_1^-)]
+[s(\theta_1^+) + s(\theta_1^-)],
$$
$$
c_y(\theta_1)=-[d(\theta_1^+) - d(\theta_1^-)]
+[s(\theta_1^+) + s(\theta_1^-)],
$$
where $\theta_1^{\pm} = \theta_1 \pm \delta /2$.
For odd twin-twin interfaces, for which $\psi_{d1}=\psi_{d2}=\psi_d$,
$$
c_x(\theta_1)=+[d(\theta_1^+) + d(\theta_1^-)]
+[s(\theta_1^+) - s(\theta_1^-)],
$$
$$
c_y(\theta_1)=-[d(\theta_1^+) + d(\theta_1^-)]
+[s(\theta_1^+) - s(\theta_1^-)].
$$

The above discussion (which assumed that the entire crystal
was made up of a single [$\,\overline 1$10] macrodomain)
can be extended to
the case where both [110] and [$\,\overline 1$10]
macrodomains are present.  By assuming that the [110] and
[$\,\overline 1$10] twin boundaries are orthogonal,
the expressions for $c_x$ and $c_y$ for the
[110] macrodomain can be shown to be identical
to those for the [$\,\overline 1$10] macrodomain,
except that $\delta$
is replaced by its negative.  Thus, if an average over
equally weighted macrodomains is taken, and the
phase of $\psi_d$ is assumed to be the same in both
types of macrodomains, the terms in $c_x$ and $c_y$
which are odd in $\delta$ drop out, and results
characteristic of perfect tetragonal symmetry are
obtained.  For even twin boundaries, Josephson currents
with $s$-wave symmetry are obtained, i.e.
$$
{\overline c}_x(\theta_1) = {\overline c}_y(\theta_1) =
s(\theta_1 + \delta /2) + s(-\theta_1 + \delta /2)
$$
where an overbar on $c$ indicates that an average over
the two macrodomains has been taken (in addition to the
previous average over the two twin types for each
macrodomain).  For odd twin boundaries, currents with
$d_{x^2 - y^2}$-wave symmetry are obtained, i.e.
$$
{\overline c}_x(\theta_1) = -{\overline c}_y(\theta_1) =
d(\theta_1 + \delta /2) + d(-\theta_1 + \delta /2).
$$
On the other hand, if $\psi_d$ changes sign on going from
the [110] macrodomain to the [$\overline 1$10]
macrodomain, then even twin boundaries yield $d_{xy}$-wave
currents with
$$
{\overline c}_x(\theta_1) = -{\overline c}_y(\theta_1) =
d(\theta_1 + \delta /2) - d(-\theta_1 + \delta /2).
$$
while odd twin boundaries yield $L_z$-wave (where
$L_z$ is the z-component of an angular momentum) currents with
$$
{\overline c}_x(\theta_1) = {\overline c}_y(\theta_1) =
s(\theta_1 + \delta /2) - s(-\theta_1 + \delta /2).
$$

Now suppose that the basal-plane Josephson currents in
YBCO, when determined on samples in which all twins and
macrodomains are present with equal weights, are
characteristic of $d_{x^2 - y^2}$-wave symmetry, as is suggested by
experiment \protect\cite{wol93,Mat95,Igu94,Tsu94}.
 From above this implies that the twin boundaries are odd
and $\psi_d$ does not change sign
between [110] and [$\,\overline 1$10] macrodomains.

It is to be emphasized that the experiments on
heavily twinned samples do not determine experimentally
whether the superconductivity in orthorhombic
YBCO is predominately $s$-wave or
predominantely $d_{x^2 - y^2}$-wave.  Above, the
function $g(\theta)$ characterizing the basal plane
Josephson currents was separated into two components,
a $d$-wave component $d(\theta)$ and an $s$-wave
component $s(\theta)$.
However, as shown above, in the case of odd twin boundaries
and $\psi_d$ not changing sign between macrodomains,
the average contribution of the function $s(\theta)$
to the Josephson currents is zero.  Thus,
the experiments measure only the
$d$-wave component $d(\theta)$, independently
of whether $d(\theta)$ is much greater than or much less
than $s(\theta)$.

Josephson tunneling experiments on untwinned single
crystals \protect\cite{wol93,Bra94} can, of course,
classify the superconductivity
as being predominantly $s$- or $d_{x^2 - y^2}$-wave.

Finally, note that calling the superconductivity predominantly
$s$- or $d$-wave according to the relative average magnitudes
of $s(\theta)$ and $d(\theta)$, while an appropriate definition
from the point of view of tunneling experiments, is only
one possible definition (and one in which the anisotropy of
the tunneling matrix elements will play a role).  Another
definition which is not necessarily equivalent could be based
on the relative magnitudes of the $s$- and $d$-wave contributions
to the gap function.

{\it Josephson currents in the $c$ direction.}
Symmetry considerations show that the $c$-axis Josephson
currents for twins 1 and 2 (e.g. see Fig. 1) have
the form
$$
j_{z\epsilon} = c_z(-1)^{\epsilon}
(\psi_{d\epsilon}^{\ast}\psi_{Pb}
-\psi_{d\epsilon}\psi_{Pb}^{\ast}).
$$
For odd twin boundaries, $\psi_{d1}=\psi_{d2}$,
and the Josephson current averages to zero when averaging over
an equal weighting of twins 1 and 2.  However, in the $c$-axis
measurements of Josephson tunneling \protect\cite{Sun94}, the
Josephson current is found not to average to zero.  Thus, these
experiments would seem to imply that the twin boundaries are
even, in contradiction to the result implied by the above
interpretation of the basal-plane Josephson current experiments.

{\it Conclusions.}
Josephson tunneling experiments measuring Josephson currents
from an orthorhombic superconductor such as YBCO into an
$s$-wave superconductor such as Pb, and carried out on
samples in which all twins and macrodomains have equal
weights, experimentally determine the twin-boundary reflection
symmetry of the superconducting state.
In the case of YBCO,
the basal-plane experiments \protect\cite
{wol93,Igu94}
characteristic of $d$-wave symmetry suggest
that the superconducting state has odd reflection
symmetry with respect to the twinning plane, whereas
a $c$-axis Josephson experiment \protect\cite{Sun94}
suggests that the superconducting state has even
twinning-plane symmetry.

A similar analysis carried out for YBCO to YBCO
Josephson tunneling
shows that here also the experiments on twinned crystals
experimentally determine
the twin boundary symmetry; the results of
\protect\cite{wol93,Mat95,Tsu94} suggest that the twin boundaries have
even symmetry, in contrast to the results of
\protect\cite{Cha94} which suggest even-symmetry
twin boundaries.

The above interpretation assumes that, in twinned crystals,
all twins and macrodomains are present with equal weights.
If this is not true, the interpretation is not valid.  For
example, if, in the experiment of Sun {\it et al.}
\protect\cite{Sun94}, twins of type 1 occupied an area of the
Josephson junction which was significantly larger than that
occupied by twins of type 2, the assumption of
odd symmetry twin boundaries would be acceptable
since there would then be only a partial cancellation
of the Josephson currents, and a nonzero
Josephson current would be observed.  The results of the
$c$-axis Josephson experiments and the basal-plane
Josephson experiments could then be reconciled.
Since this question of the weighting of the different
twins and macrodomains is of crucial importance
to the interpretation, it would be of interest to try
to determine this weighting experimentally, for example by
microstructural studies of the Josephson junctions used
for both the $c$-axis and the basal-plane experiments.

Another important conclusion is that, contrary to
currently accepted ideas, Josephson tunneling experiments
on orthorhombic YBCO crystals containing all twins
and macrodomains with equal weights can not directly
experimentally determine whether the superconductivity is
predominately $d$- or $s$-wave (but can, as described
above, experimentally determine the nature of the twin boundaries).
Our direct experimental knowledge of the order parameter
symmetry as determined by Josephson experiments thus
comes from experiments performed on untwinned
single crystals \protect\cite{Bra94,wol93}.

\section*{Acknowledgements}

Stimulating discussions with R.C. Dynes and
J. Luettmer-Strathmann are acknowledged, as is the
hospitality of P. Nozi\`eres and the theory group of the
Institut Laue-Langevin where part of the work
was carried out.  This work was supported by
the Natural Sciences and Engineering Research Council of
Canada.

\newpage
\begin{center}
{\bf Figure Captions}
\end{center}
\vspace{1 cm}
\begin{description}
\item[Fig.~1] (a) Shows a [$\,\overline 1$10]
twin boundary(DB), the two
twins with the directions of their $a$ and $b$ axes, and
the interface between YBCO and Pb (ABC) with its
normal {\bf n}. (b)Definition of the angles between
{\bf a$_1$, a$_2$}, and {\bf n}.
\end{description}

\end{document}